\documentclass[12pt]{article}
\usepackage{a4wide}

\usepackage{epsfig}
\usepackage{subfigure}
\usepackage{graphicx}
\usepackage{amsmath}
\usepackage{cite}

\usepackage[usenames]{color}

\usepackage[normalem]{ulem}
\newcommand{\Tr}{{\rm Tr}}

\usepackage{latexsym}
\usepackage{array}
\usepackage{verbatim}

\usepackage{rotating}
\usepackage{color}

\usepackage[utf8]{inputenc}

\def\lsim{\mathrel{\raise.3ex\hbox{$<$\kern-.75em\lower1ex\hbox{$\sim$}}}}
\def\gsim{\mathrel{\raise.3ex\hbox{$>$\kern-.75em\lower1ex\hbox{$\sim$}}}}

\newcommand{\Lumint}{{\cal L}_{\rm int}}

\usepackage{amssymb}

\allowdisplaybreaks

\begin{document}
\begin{titlepage}
\begin{center}

{\large \bf {High-precision limits on $W$-$W'$ and $Z$-$Z'$ mixing
from diboson production using the full LHC Run~2 ATLAS data set}}

\vskip 1cm

A.~A. Pankov,$^{a,b,c,}$\footnote{E-mail: pankov@ictp.it}
P. Osland,$^{d,}$\footnote{E-mail: Per.Osland@uib.no}
I.~A. Serenkova$^{a,}$\footnote{E-mail: Inna.Serenkova@cern.ch} and
V.~A. Bednyakov$^{c,}$\footnote{E-mail: bedny@jinr.ru}

\vspace{1.0cm}

$^{a}$The Abdus Salam ICTP
Affiliated Centre, Technical University of Gomel, \\
246746 Gomel, Belarus, \\
$^{b}$Institute for Nuclear Problems, Belarusian
 State University, 220030 Minsk, Belarus, \\
$^{c}$Joint Institute  for Nuclear Research, Dubna 141980 Russia,\\
$^{d}$Department of Physics and Technology, University of Bergen, \\
Postboks 7803, N-5020  Bergen, Norway

\end{center}

\vskip 3cm

\begin{abstract}
The full ATLAS Run~2 data set with time-integrated luminosity of 139 fb$^{-1}$ in
the diboson channels in hadronic final states is used to probe a simple model
with an extended gauge sector (EGM), proposed by Altarelli et al., and  often taken
as a convenient benchmark by experimentalists. This model accommodates new
charged $W'$ and neutral $Z'$ vector bosons with modified
trilinear Standard Model gauge couplings, decaying into
electroweak gauge boson pairs $WZ$ or $WW$, where $W$/$Z$ decay hadronically.
 Exclusion limits at the 95\% C.L. on the
$Z'$ and $W'$ resonance production cross section times branching
ratio to electroweak gauge boson pairs in
the mass range of $\sim$ 1 -- 5 TeV  are here converted to
constraints on $W$-$W'$ and $Z$-$Z'$ mixing parameters and
masses for the EGM.
We present exclusion regions on the
parameter space of the $W'$ and $Z'$ by using the full Run~2
data set comprised of $pp$ collisions at $\sqrt{s}=13$ TeV and
recorded by the ATLAS detector at the CERN LHC. The obtained
exclusion regions are significantly extended compared to
those obtained from the previous analysis performed with Tevatron
data as well as with LHC data collected at 7 and 8 TeV in Run~1
and are the most stringent bounds to date.
\end{abstract}

\end{titlepage}

%%%%%%%%%%%%%%%%%%%%%%%%%%%%%%%%%%%%%%%%%%%%%%%%%%%%%%%%%%%%
\section{Introduction} \label{sec:I}
%%%%%%%%%%%%%%%%%%%%%%%%%%%%%%%%%%%%%%%%%%%%%%%%%%%%%%%%%%%%

One of the main aims of the physics programme at the Large Hadron Collider (LHC) is to search
for new  phenomena that become visible in high-energy proton-proton collisions.
A possible signature of such new phenomena would be the production of a heavy resonance with
its subsequent decay into a final state consisting of a pair of fermions or vector bosons.
Many new physics scenarios beyond the Standard Model (SM) predict such a signal. Possible candidates
are charged and neutral heavy gauge bosons.
In the simplest models these particles are considered copies of the SM $W$ and $Z$
bosons and are commonly referred to as $W'$ and $Z'$ bosons \cite{Tanabashi:2018oca}.
In the Sequential Standard Model (SSM) \cite{Altarelli:1989ff} the $W'_{\rm SSM }$ and  $Z'_{\rm SSM }$ bosons have couplings to fermions that are identical to those of the SM $W$ and $Z$
bosons, but for which the trilinear couplings $W'WZ$ and $Z'WW$ are absent.
The SSM has been used as a reference for experimental $W'$ and $Z'$
boson searches for decades, the results can be re-interpreted in the context of
other models of new physics, and it is useful for comparing the
sensitivity of different experiments.

At the LHC, such heavy $W'$ and $Z'$
bosons could be observed through their single production as $s$-channel
resonances with subsequent leptonic decays
\begin{equation}
pp\to W^\prime X \to \ell\nu X, \label{proclept}
\end{equation}
and
\begin{equation}
pp\to Z^\prime X \to \ell^+\ell^- X, \label{procleptt}
\end{equation}
respectively, where in what follows, $\ell=e,\mu$ unless otherwise stated.
The production of $W'$ and $Z'$  bosons at hadron colliders is expected to be dominated by the process
$q\bar{q}^\prime/q\bar{q}\to W'/Z'$.
Leptonic final states provide a low-background and efficient experimental signature that
results in  excellent sensitivity to new phenomena at the LHC.
Specifically, these processes (\ref{proclept}) and (\ref{procleptt}) offer the simplest event topology for the
discovery of $W^\prime$ and $Z^\prime$ with a large production rate and a clean
experimental signature. These channels are among the most promising discoveries at the
LHC \cite{Aad:2019fac, CMS:2019tbu, Zucchetta:2019afp,CMS:2018wsn, Aad:2019wvl}.
There have also been many theoretical studies of $W^\prime$ and $Z^\prime$
boson searches
at the high energy hadron colliders (see, e.g. \cite{Tanabashi:2018oca,Langacker:1984dc,Langacker:1991pg,Schmaltz:2010xr,Grojean:2011vu,Jezo:2012rm,Cao:2012ng,Dobrescu:2015yba,
Chavez:2019yal,Langacker:2008yv,Erler:2009jh,Hewett:1988xc,Leike:1998wr,Dittmar:2003ir,Osland:2009tn,Godfrey:2013eta,Andreev:2014fwa,Gulov:2018zij,Bandyopadhyay:2018cwu}).

The data we consider were collected with the ATLAS and CMS detectors during the 2015--2018 running period of the LHC,
referred to as Run 2 and corresponding to time-integrated luminosity of 139~fb$^{-1}$.
The ATLAS  experiment has presented the first search for dilepton resonances  based on the full Run 2 data set \cite{Aad:2019fac,  Aad:2019wvl} and set limits on the $W'$ and $Z'$
production cross sections times branching fraction in the processes
(\ref{proclept}) and (\ref{procleptt}), $\sigma(pp\to W'X)\times \text{BR}(W'\to \ell\nu)$ and $\sigma(pp\to Z'X)\times \text{BR}(Z'\to \ell^+\ell^-)$, respectively, for
$M_{W'}$ and  $M_{Z'}$ in the 0.15 TeV -- 7 TeV and  0.25 TeV -- 6 TeV ranges, correspondingly. Recently, similar searches have also been presented by the CMS Collaboration  using  140~fb$^{-1}$ of data recorded at $\sqrt{s}=13$ TeV \cite{CMS:2019tbu}. The most stringent limits on the mass of $W'_{\rm SSM}$ and $Z'_{\rm SSM}$ bosons to date come from the searches in respectively,
(\ref{proclept}) and (\ref{procleptt}) processes
by the ATLAS and CMS collaborations using data taken at $\sqrt{s} = 13$
TeV in Run 2 and set a 95\% confidence level (CL) lower limit on the $W'_{\rm SSM}$
mass of 6.0~TeV  \cite{Aad:2019wvl} and $\sim$ 5.2 TeV for $Z'_{\rm SSM}$ \cite{Aad:2019fac,CMS:2019tbu}.

Alternative $W'$ and $Z'$ search channels are the diboson reactions
\begin{equation}
{ pp\to W^\prime X\to WZ X,} \label{procWZ}
\end{equation}
and
\begin{equation}
{ pp\to Z^\prime X\to WW X.} \label{procWW}
\end{equation}
The study of gauge boson pair production offers a powerful test of the
spontaneously broken gauge symmetry of the SM and can be used as a
probe for new phenomena beyond the SM.

Heavy resonances that can decay to gauge boson pairs are
predicted in many scenarios of new physics, including extended
gauge models (EGM) \cite{Altarelli:1989ff,Eichten:1984eu}, models
of warped extra dimensions~\cite{Randall:1999ee,Davoudiasl:2000wi}, technicolour
models~\cite{Lane:2002sm,Eichten:2007sx} associated with the existence
of technirho and other technimesons, more generic composite
Higgs models \cite{Agashe:2004rs, Giudice:2007fh}, and the heavy
vector-triplet (HVT) model \cite{Pappadopulo:2014qza}, which generalises a large number of
models that predict spin-1 charged ($W'$) and neutral ($Z'$)
resonances etc.
 Searches for exotic heavy particles that decay into $WZ$ or $WW$ pairs are complementary
to searches in the leptonic channels $\ell\nu$ and $\ell^+\ell^-$ of the processes
(\ref{proclept})  and (\ref{procleptt}). Moreover, there are models in which
new gauge boson couplings to SM fermions are suppressed, giving rise to a
fermiophobic $W'$ and $Z'$ with an enhanced coupling to electroweak gauge bosons~\cite{Tanabashi:2018oca, He:2007ge}. It is therefore important to
search for $W'$ and $Z'$ bosons also in the $WZ$ and $WW$ final states.

The paper is organized as follows.
In Sect.~\ref{sect:framework} we present the theoretical framework, then, in
Sect.~\ref{sect:production} we summarize the relevant cross sections
for processes (\ref{procWZ}) and (\ref{procWW})  in the narrow width approximation (NWA) to the EGM. Next, in Sect.~\ref{sect:next} we discuss the relevant $W'$ and $Z'$ branching ratios.
In Sect.~\ref{sect:analysis}, we present an analysis of bounds on $W$-$W'$ and $Z$-$Z'$ mixing from constraints on diboson production in the context of the EGM, employing the most recent measurements recorded by the ATLAS (36.7 fb$^{-1}$ and 139 fb$^{-1}$) detector  \cite{Aaboud:2017eta,Aad:2019fbh} at the LHC. Then,  we show the resulting constraints on the $M_{W'}-\xi_{W\text{-}W'}$ and $M_{Z'}-\xi_{Z\text{-}Z'}$ parameter space obtained from the diboson processes, (\ref{procWZ}) and (\ref{procWW}). Further,  we
collect and compare the indirect constraints obtained from
electroweak precision data, direct search constraints derived from the Tevatron and
at the LHC in Run~1 and Run~2 data.
Sect.~\ref{sect:conclusions} presents some concluding remarks.

%%%%%%%%%%%%%%%%%%%%%%%%%%%%%%%%%%%%%%%%%%%%%%%%%
\section{Framework}
\label{sect:framework}
\setcounter{equation}{0}
%%%%%%%%%%%%%%%%%%%%%%%%%%%%%%%%%%%%%%%%%%%%%%%%%

Because of the large variety of models which predict new heavy charged and neutral
gauge bosons, after a discovery of
signatures associated to a new boson, detailed studies must be
carried out to distinguish between these models and to determine
whether the boson belongs to one of the theoretically motivated
models such as, e.g.\ EGM or some other model. Following
the traditions of direct searches at hadron colliders, such
studies are based on the model first proposed in
Ref.~\cite{Altarelli:1989ff}.

As mentioned above, in the SSM, the coupling constants of the
$W'$ and $Z'$ bosons with SM fermions are identical to the
corresponding SM couplings, while the $W'$ and $Z'$ couplings to, respectively,  $WZ$ and $WW$
vanish, $g_{W'WZ}=g_{Z'WW}=0$. Such a suppression may arise in an EGM in a natural
manner: if the new gauge bosons and the SM ones belong to
different gauge groups, vertices such as $W'WZ$ and $Z'WW$ do not arise. They
can only occur after symmetry breaking due to mixing of the gauge
eigenstates. Triple gauge boson couplings (such as $W'WZ$ and $Z'WW$) as well
as the vector-vector-scalar couplings (like $W'WH$ and $Z'ZH$) arise from the
symmetry breaking and may contribute to the $W'$ and $Z'$ decays, respectively.
The vertices are then suppressed by a factor of the order of
$(M_W/M_{V'})^2$, where $V^{\prime}$  represents  a $W^{\prime}$ or a $Z^{\prime}$ boson.

In an EGM \cite{Altarelli:1989ff}, the trilinear gauge boson couplings are modified by mixing
factors
\begin{equation} \label{Eq:define-xi}
\xi_{V\text{-}V'}={\cal C} \times (M_W/M_{V'})^2,
\end{equation}
where ${\cal C}$ is a scaling constant that sets the coupling
strength. Specifically,
in an EGM the standard-model trilinear gauge boson coupling
strength $g_{WWZ}$ ({$=e\cot\theta_W$}), is replaced by
$g_{W'WZ}=\xi_{W\text{-}W^\prime}\cdot g_{WWZ}$ in the $WZ$ channel and  $g_{Z'WW}=\xi_{Z\text{-}Z^\prime}\cdot g_{WWZ}$  in the $WW$ channel. Following the parametrization of the
trilinear gauge boson couplings $W'WZ$ and $Z'WW$ presented in \cite{Aaltonen:2010ws}
for the analysis and interpretation of the
CDF data on  $p\bar{p}\to W'X\to WZX$  and $p\bar{p}\to Z'X\to W^+W^-X$, expressed in terms of
two free parameters,\footnote{Such a $W'$ and $Z'$, described in terms of
two parameters, are here referred to as the EGM bosons.} $\xi_{W\text{-}W^\prime}$ ($\xi_{Z\text{-}Z^\prime}$) and
$M_{W'}$ ($M_{Z'}$), we will set  $W'$ ($Z'$) limits as
functions of the mass $M_{W'}$ ($M_{Z'}$) and mixing factor $\xi_{W\text{-}W^\prime}$ ($\xi_{Z\text{-}Z^\prime}$)  by using
the ATLAS resonant diboson production data \cite{Aaboud:2017eta,Aad:2019fbh}
collected at a center of mass energy of $\sqrt{s}=13$ TeV, taking into account the partial and full Run~2 data sets with time-integrated luminosity of 36.7 fb$^{-1}$ and 139 fb$^{-1}$, respectively.
The presented analysis in the EGM with two free parameters is more
general than the previous ones where the only parameter is the
$V'$ mass. As for the SSM, one has $V'_{\rm SSM}\equiv V'_{\rm EGM}(\xi_{V\text{-}V'}=0)$.

The parametrization of boson mixing introduced by Altarelli et al. \cite{Altarelli:1989ff}, though being simplified, has a well-motivated theoretical basis. To be specific,  we briefly consider $Z^0$--$Z^{0\prime}$ mixing within the framework of  models with extended gauge sector such as the $E_6$ models, the LR model and SSM (see, e.g.
\cite{Langacker:2008yv,Erler:2009jh,Hewett:1988xc,Leike:1998wr,Langacker:1991pg}).

The physical (mass eigenstates) $Z$ and $Z^\prime$ are admixtures of the weak eigenstates $Z^0$ of $SU(2)\times U(1)$ and $Z^{0\prime}$ of the extra $U(1)'$, respectively.   The mass eigenstates, $Z$ and $Z^\prime$ are obtained by a rotation of the fields $Z^0$ and $Z^{0\prime}$:
\label{Eq:Z12-couplings}
\begin{eqnarray}
&Z& = Z^0\cos\phi + Z^{0\prime}\sin\phi\;, \label{z} \\
&Z^\prime& = -Z^0\sin\phi + Z^{0\prime}\cos\phi\;. \label{zprime}
\end{eqnarray}

For each type of $Z'$ boson, defined by each set of gauge couplings, there are three classes of models, which differ in the assumptions concerning the quantum numbers of the Higgs fields which generate the $Z$-boson mass matrix \cite{Langacker:2008yv,Erler:2009jh,Langacker:1991pg}. In each case there is a relation between the $Z^0$-$Z^{0\prime}$ mixing angle $\phi$ and the two mass eigenvalues $M_Z$ and $M_{Z'}$ which can be written as \cite{Langacker:1984dc,Langacker:2008yv}:
\begin{equation} \label{phi}
\tan^2\phi={\frac{M_{Z^0}^2-M_Z^2}{M_{Z'}^2-M_{Z^0}^2}}\;,
\end{equation}
where ${M_{Z^0}}$ is the mass of the $Z$ boson in the absence of mixing, i.e., for
$\phi=0$. The mixing angle $\phi$ will play an important role in our analysis. Such mixing effects reflect the underlying gauge symmetry and/or the Higgs sector of the model:
\begin{itemize}
\item[(i)]
The least constrained ($\rho_0$ free) model makes no assumption concerning the
Higgs sector. It allows arbitrary $SU(2)$ representations
for the Higgs fields, and is the analog of allowing  $\rho_0\neq 1$  in
the $SU(2)\times U(1)$ model. In this case $M_Z$, $M_{Z'}$ and $\phi$ are all free parameters.
\item[(ii)]
If one assume that all $SU(2)$ breaking is due to Higgs doublets and singlets  ($\rho_0=1$ model), there are only two free parameters, which we identify as $\phi$ and $M_{Z^\prime}$, and we will adopt this parametrization throughout the paper, specifically for the EGM case.
\item[(iii)]
Finally, in specific models one specifies not only the $SU(2)$ assignments
but the $U(1)^\prime$ assignments of the Higgs fields.
Since the same Higgs multiplets generate both $M_Z$ and $\phi$, one
has an additional constraint.    To a good approximation,
for $M_Z\ll M_{Z'}$, in specific ``minimal-Higgs models'', one has an
additional constraint \cite{Langacker:1984dc}
\begin{equation}\label{phi0}
\phi\simeq -s^2_\mathrm{W}\
\frac{\sum_{i}\langle\Phi_i\rangle{}^2I^i_{3L}Q^{\prime}_i}
{\sum_{i}\langle\Phi_i\rangle^2(I^i_{3L})^2} =
P\,{\frac{\displaystyle M^2_Z}{\displaystyle M^2_{Z'}}},
\end{equation}
where $s_\mathrm{W}$ is the sine of the electroweak  angle.
In these models $\phi$ and $M_{Z'}$ are not independent and
there is only one (e.g., $M_{Z'}$) free parameter. This parametrization corresponds to the expression of the mixing factor presented in Eq.~(\ref{Eq:define-xi}).
Furthermore,
$\langle\Phi_i\rangle$ are the Higgs (doublet) vacuum expectation values
spontaneously breaking the symmetry, and $Q^\prime_i$  are their
charges with respect to the additional $U(1)'$. In
these models the same Higgs multiplets are responsible for both
generation of the mass $M_Z$ and for the strength of the
$Z^0$--$Z^{0\prime}$ mixing. Thus $P$ is a model-dependent
constant.
\end{itemize}

This mixing between $Z^0$ and $Z^{0\prime}$ will induce a change in couplings of the two bosons to fermions.
An important property of the models
under consideration is that the gauge eigenstate $Z^{0\prime}$ does
not couple to the $W^+W^-$ pair since it is neutral under
$SU(2)$. Therefore the $W$-pair production is sensitive to a
$Z^\prime$ only in the case of a non-zero $Z^0$--$Z^{0\prime}$ mixing.
From (\ref{z}) and (\ref{zprime}), one obtains:
\begin{subequations}
\begin{eqnarray}
&& g_{WWZ}=\cos\phi\;g_{WWZ^{0}}\;, \label{WWZ1} \\
&& g_{WWZ'}=-\sin\phi\; g_{WWZ^{0}}\;,\label{WWZ2}
\end{eqnarray}
\end{subequations}
where $g_{WWZ^{0}}=e\cot\theta_W$. Also,
$g_{WW\gamma}=e$.

In many extended gauge models, while the couplings to fermions are
not much different from those of the SM, the $Z'WW$ coupling is
substantially suppressed with respect to that of the SM. In fact,
in the extended gauge models  the SM trilinear gauge boson
coupling strength, $g_{WWZ^{0}}$, is replaced by $g_{WWZ^{0}} \rightarrow
\xi\cdot g_{WWZ^{0}}$, where $\xi \equiv \vert\sin\phi\vert$
(see Eq.~(\ref{WWZ2})) is the mixing factor\footnote{For weak mixing, $\xi\simeq|\phi|$, and is therefore often referred to as a mixing ``angle''.}.
We will set cross section
limits on such $Z'$ as functions of the mass $M_{Z'}$ and $\xi$.

Previous analyses of the $Z$-$Z'$ and $W$-$W'$ mixing \cite{Osland:2017ema,Bobovnikov:2018fwt,Serenkova:2019zav}\footnote{Strictly speaking, ``$Z$-$Z'$ mixing'' should be referred to as ``$Z^0$-$Z^{0\prime}$ mixing'' and similarly for ``$W$-$W'$ mixing''.}
were carried out using the diboson production data set corresponding to the time-integrated luminosity of $\sim$ 36 fb$^{-1}$  collected in 2015 and 2016 with the ATLAS and CMS collaborations at $\sqrt{s}=$ 13~TeV where electroweak $Z$ and $W$ gauge bosons decay into the semileptonic channel \cite{Aaboud:2017fgj} or into the dijet final state \cite{Sirunyan:2017acf}.
The results of the present analysis benefit from the increased size of the data sample corresponding to an integrated luminosity of  139 fb$^{-1}$
recorded by the ATLAS detector in Run~2 \cite{Aad:2019fbh} which is almost four times larger than what was available for the previous study in the semileptonic final state\footnote{In the current analysis, we utilize the full Run~2 ATLAS
data set on diboson resonance production \cite{Aad:2019fbh}, rather than that of CMS, as the latter one is unavailable so far.}. In addition,
further improvement in placing limits on the $W'$ and $Z'$ mass and $W$-$W'$ and $Z$-$Z'$  mixing parameters can be achieved in fully-hadronic $WZ/WW\to qqqq$ final states\footnote{To simplify notation, antiparticles are denoted by the same symbol as the corresponding particles.} using the novel reconstruction and analysis techniques of a diboson system with pairs of large-radius jets.
Indeed,  the $W$ and $Z$ bosons produced in the decay of TeV-scale resonances are highly energetic (``boosted'') so that their decay products are merged into a single large-radius jet, and are therefore reconstructed experimentally as a single large-radius-parameter jet and accordingly, interpreted as a two-jet final state.  The signature of such heavy resonance decays is thus a resonant structure in the dijet invariant mass spectrum. This novel technique allows to improve background estimation and the signal extraction procedure, resulting in higher sensitivity of the analysis.

The properties of possible $W'$ and $Z'$ bosons are also constrained by measurements
of electroweak (EW) processes at low energies, i.e., at energies much below the masses of new
charged and neutral gauge bosons.
Such bounds on the $W$-$W'$ ($Z$-$Z'$) mixing are mostly due to the
deviation in $W$ ($Z$) properties compared to the SM predictions. These
measurements show that the mixing angles $\xi_{W\text{-}W^\prime}$ and $\xi_{Z\text{-}Z^\prime}$  between the gauge
eigenstates must be smaller than about $10^{-2}$ and $2.6\cdot 10^{-3}$, respectively \cite{Tanabashi:2018oca,Erler:2009jh}.

In this work, we derive bounds on the possible new spin-1
resonances ($W^\prime$/$Z^\prime$) within the EGM framework, from the
full ATLAS Run~2 data set on $WZ$/$WW$ pair production with time-integrated luminosity of 139 fb$^{-1}$ \cite{Aad:2019fbh}.  The search was conducted for a $W'$/ $Z'$
resonance decaying into a $WZ$/$WW$ boson pair, where the $W$ and $Z$ bosons
decay hadronically. We present results as constraints on the relevant
$W$-$W'$ ($Z$-$Z'$)
mixing angle,  $\xi_{W\text{-}W^\prime}$ ($\xi_{Z\text{-}Z^\prime}$),  and on the mass $M_{W'}$ ($M_{Z'}$) and display the
combined allowed parameter space for the benchmark $W'$ ($Z'$) bosons,
showing also indirect constraints from electroweak precision data, previous direct search constraints from the Tevatron and from the LHC with 7 and 8~TeV in Run~1 as well as those obtained from the LHC at 13 TeV with a partial ATLAS Run~2 data set with time integrated luminosity of 36.7 fb$^{-1}$ \cite{Aaboud:2017eta}  in the fully hadronic ($qqqq$) final states.

Let us here comment on possible mechanisms that might generate the $V$-$V'$ mixing. Within a UV-complete theory, mixing could enter the trilinear coupling via the kinetic terms,
\begin{equation}
-\frac{1}{2}\Tr\left[V_{\mu\nu}^\dagger V^{\mu\nu}\right], \quad \text{with }
V_{\mu\nu}=[D_\mu,D_\nu],
\end{equation}
where the covariant derivative includes the heavier gauge field, $V'_\mu$, schematically
\begin{equation}
D_\mu=D_\mu^\text{SM}+g'V'_\mu.
\end{equation}
An off-diagonal term in the mass-squared matrix would lead to mixing as given by Eq.~(\ref{phi}).
On the other hand, mixing could be a loop effect. While such examples of mechanisms do not offer much insight on the magnitude of the mixing, they would allow for an interpretation of an observed signal.

%%%%%%%%%%%%%%%%%%%%%%%%%%%%%%%%%%%%%%%%%%%%%%%%%
\section{Resonant diboson production in $pp$ collision}
\label{sect:production}
\setcounter{equation}{0}
%%%%%%%%%%%%%%%%%%%%%%%%%%%%%%%%%%%%%%%%%%%%%%%%%

At lowest order in the EGM, $W'$ production and decay into $WZ$ in
proton-proton collisions occurs through quark-antiquark
interactions in the $s$-channel.
The  cross section of process (\ref{procWZ})  can at the LHC be
observed through resonant pair production of gauge bosons $WZ$.
Using the NWA, one can factorize the
process (\ref{procWZ}) into the $W'$ production and the $W'$
decay,
\begin{equation}
\sigma(pp\to W' X\to WZX)  = \sigma(pp\to W'X) \times \text{BR}(W' \to
WZ)\;. \label{TotCr}
\end{equation}
Here, $\sigma(pp\to W' X)$ is the  total (theoretical) $W'$
production cross section and
$\text{BR}(W' \to WZ)=\Gamma_{W'}^{WZ}/\Gamma_{W'}$ with
$\Gamma_{W'}$ the total width of $W'$. ``Narrow'' refers to the assumption that the natural width of a resonance is smaller than the typical experimental resolution of 5\% of its mass \cite{Aaboud:2016okv,Sirunyan:2017nrt}, which is true for a large fraction of the parameter space of the reference EGM model.

Likewise, $Z'$ production and decay into $WW$ can be
observed through resonant pair production of charged gauge bosons $WW$.
In the NWA, one can write down the cross section of process (\ref{procWW}) as follows:
\begin{equation}
\sigma(pp\to Z' X\to WWX)  = \sigma(pp\to Z'X) \times \text{BR}(Z' \to
WW)\;. \label{TotCrWW}
\end{equation}
Here, $\sigma(pp\to Z'X)$ is the  total (theoretical) $Z'$
production cross section and
$\text{BR}(Z' \to WW)=\Gamma_{Z'}^{WW}/\Gamma_{Z'}$ with
$\Gamma_{Z'}$ the total width of $Z'$.

%%%%%%%%%%%%%%%%%%%%%%%%%%%%%%%%%%%%%%%%%%%%%%%%%%%%%%%%%%%%
\section{$W^\prime$ and $Z^\prime$ Branching Ratios}
\label{sect:next}
\setcounter{equation}{0}
%%%%%%%%%%%%%%%%%%%%%%%%%%%%%%%%%%%%%%%%%%%%%%%%%%%%%%%%%%%%
We shall here review the decay modes of $W'$ and $Z'$, with a focus on their branching ratios to $WZ$ and $WW$, respectively.

%%%%%%%%%%%%%%%%%%%%%%%%%%%%%%%%%%%%%%%%%%%%%%%%%
\subsection{$W'\to WZ$}
\label{WprimetoWZ}
%%%%%%%%%%%%%%%%%%%%%%%%%%%%%%%%%%%%%%%%%%%%%%%%%

In the EGM the $W'$ bosons can decay into SM fermions, gauge
bosons ($WZ$), or a pair of the charged SM $W$ boson and the
Higgs boson $H$. In the calculation
of the total width $\Gamma_{W'}$ we consider the following
channels: $W'\to f{\bar{f}}^\prime$, $WZ$, and $WH$, where
$f$ are SM fermions ($f=\ell,\nu,q$) \footnote{Here, the $\ell$ includes $\tau$ leptons.}.
Only left-handed neutrinos are considered, while
possible right-handed neutrinos are assumed to be kinematically
unavailable as final states. Also, throughout the paper we shall
ignore the couplings of the $W'$ to other beyond-SM particles such as
SUSY partners and exotic fermions in the theory.
The presence of such channels
would increase the width of the $W'$ and hence lower the
branching ratio into a $WZ$ pair. As a result, the total decay
width of the $W^\prime$ boson is taken to be
\begin{equation}\label{gamma}
\Gamma_{W'} = \sum_f \Gamma_{W'}^{f\bar{f}'} + \Gamma_{W'}^{WZ} +
\Gamma_{W'}^{WH}.
\end{equation}
The fermion contribution, $\sum_{ff'} \Gamma_{W'}^{f\bar f'}$, would depend on the number
$n_g$ of generations of
heavy exotic fermions which can contribute to the $W'$ decay without
phase space suppression. This number is model dependent too, and
introduces a phenomenological uncertainty.
The presence of the last two decay channels,  which are often
neglected at low and moderate values of $M_{W'}$, is due to
$W$-$W'$ mixing which is constrained to be tiny.
In particular, for the
range of $M_{W'}$ values below $\sim 1.0-1.5$ TeV,  the dependence
of $\Gamma_{W'}$ on the values of $\xi_{W\text{-}W^\prime}$ (within its allowed range)
induced by $\Gamma_{W'}^{WZ}$ and $\Gamma_{W'}^{WH}$ is
unimportant because $\sum_f \Gamma_{W'}^{f\bar {f'}}$ dominates
over diboson partial widths. Therefore, in this mass range, one
can approximate the total width as $\Gamma_{W'} \approx \sum_f
\Gamma_{W'}^{f\bar {f'}}=3.5\%\times M_{W'}$ \cite{Serenkova:2019zav},
 where the sum runs over SM fermions only.

For heavier $W'$ bosons, the diboson decay channels, $WZ$ and
$WH$, start to play an important role,
and we are no longer able to ignore them \cite{Serenkova:2019zav}.
To be specific, we take an approach as model-independent as
possible, and for numerical illustration show our results in two
simple scenarios. In the first scenario,
we treat the model as effectively having a negligible partial
width of $W'\to WH$ with respect to that of $W'\to WZ$, i.e.
$\Gamma_{W'}^{WH}\ll\Gamma_{W'}^{WZ}$, so that one can ignore the
former, taking $\Gamma_{W'}^{WH}\simeq 0$. In this case, numerical
results with our treatment will serve as an upper bound on the
size of the signal. The second scenario assumes that
both partial widths are comparable, $\Gamma_{W'}^{WH}\simeq
\Gamma_{W'}^{WZ}$ for heavy $M_{W'}$, as required by the
``Equivalence theorem'' \cite{Chanowitz:1985hj}.
%%%%%%%%%%%%%%%%%%%%%%%%%%%%%%%%%%%%%%%%%%%%%%%%%%%%%%%%
\begin{figure}[htb]
\refstepcounter{figure}
\label{BR-wprime}
\addtocounter{figure}{-1}
\begin{center}
\includegraphics[scale=0.6]{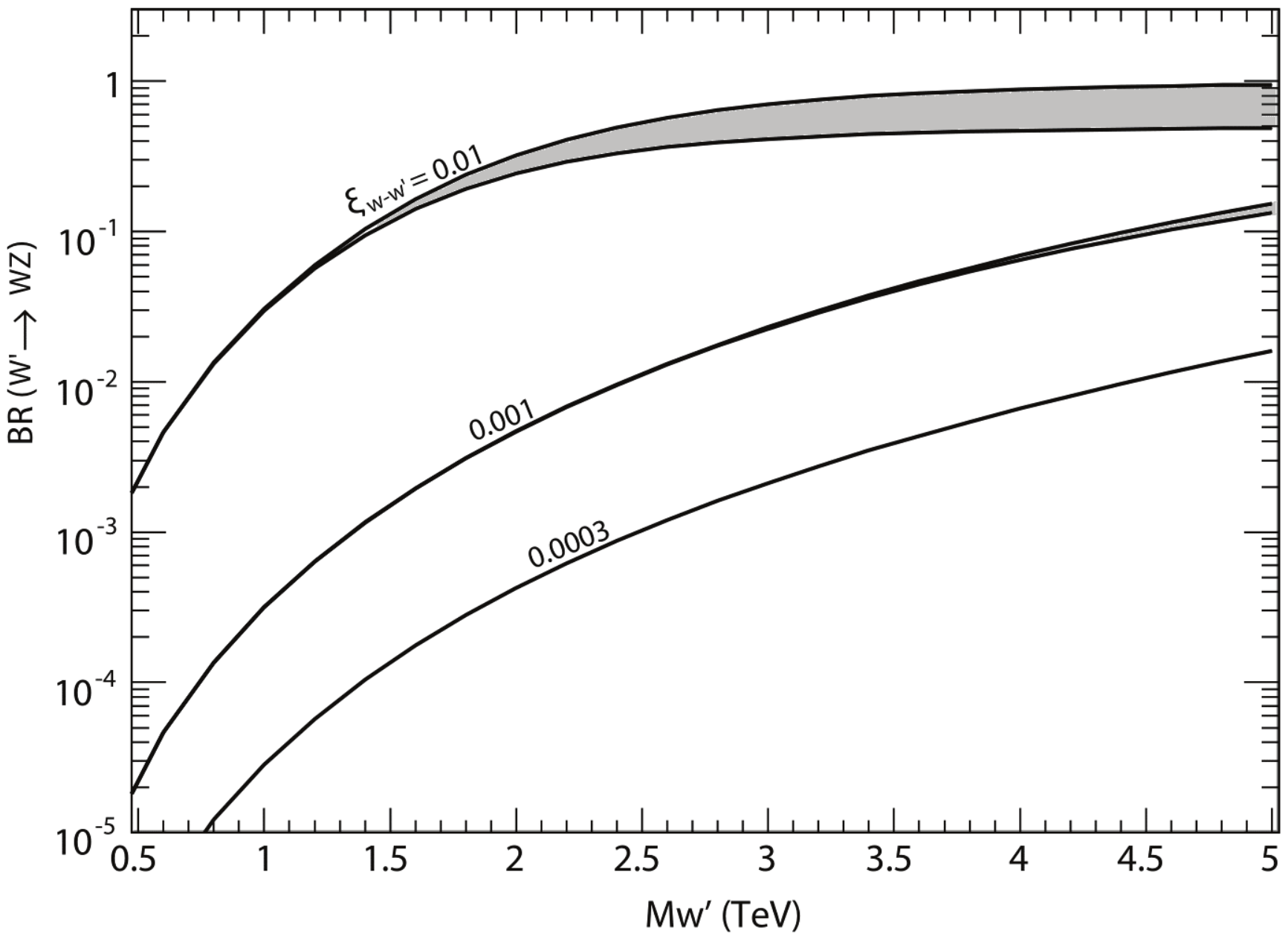}
\end{center}
\caption{Branching fraction $\text{BR}(W'\to WZ)$ (solid) vs
$M_{W'}$ in the EGM for $W$-$W'$ mixing factor
$\xi_{W\text{-}W^\prime}=3\cdot 10^{-4}$, $10^{-3}$ and $10^{-2}$.
The shaded bands represent the uncertainty resulting from the
inclusion of the $WH$ decay mode, the upper and lower bounds
correspond to the assumptions $\Gamma_{W'}^{WH}=0$ and
$\Gamma_{W'}^{WH}=\Gamma_{W'}^{WZ}$, respectively.}
\end{figure}
%%%%%%%%%%%%%%%%%%%%%%%%%%%%%%%%%%%%%%%%%%%%%%%%%%%%%%%%

In the first scenario, where $\Gamma_{W'}^{WH}=0$, for a fixed mixing
factor $\xi_{W\text{-}W^\prime}$  and at large $M_{W'}$, where $\Gamma_{W'}^{WZ}$
dominates over $\sum_f \Gamma_{W'}^{f\bar {f'}}$,  the total width
increases rapidly with the $W'$ mass because of the quintic
dependence on the $M_{W'}$ mass of the $WZ$ mode,
$\Gamma_{W'}^{WZ}\propto M_{W'}\left[{M_{W'}^4}/({M_W^2M_Z^2})\right]$,
 corresponding to the production of longitudinally polarized $W$ and $Z$ in the channel $W'\to
W_LZ_L$ \cite{Altarelli:1989ff,Serenkova:2019zav}.
In this case, the $WZ$ mode becomes dominant and
$\text{BR}(W' \to WZ)\to 1$, while the fermionic decay channels,
$\sum_f \Gamma_{W'}^{f\bar {f'}}\propto M_{W'}$, are increasingly
suppressed. However, in the second scenario with
$\Gamma_{W'}^{WH}=\Gamma_{W'}^{WZ}$, $\text{BR}(W' \to WZ)\to 0.5$
when $M_{W'}$ increases, as illustrated in
Fig.~\ref{BR-wprime}.

%%%%%%%%%%%%%%%%%%%%%%%%%%%%%%%%%%%%%%%%%%%%%%%%%
\begin{table}[htb]
\caption{$W^\prime$ branching ratios in per cent for
$\xi_{W\text{-}W^\prime}=(M_W/M_{W'})^2$.}
\begin{center}
\begin{tabular}{|l|c|c|c|c|c|}
\hline\hline
\ \ \ \ $M_{W'}$~[TeV]  & 1 & 2 & 3 & 4 & 5 \\
\hline
 BR $\to$ light quarks
 & 49.3 & 48.9 & 48.9 & 48.8 & 48.8\\
\hline
BR $\to t\bar b$
&   23.6  & 24.2 &  24.3 & 24.4 & 24.4 \\
\hline
BR $\to \ell\nu$
& 24.7  &  24.5 & 24.4 & 24.4 & 24.4 \\
\hline
BR $\to WZ+WH$
& 2.4 & 2.4 & 2.4 & 2.4 & 2.4 \\
\hline
\end{tabular}
\end{center}
\label{Tab:BR-case1}
\end{table}
%%%%%%%%%%%%%%%%%%%%%%%%%%%%%%%%%%%%%%%%%%%%%%%%%

Before closing the discussion of the diboson branching ratios, we compare them in Tables~\ref{Tab:BR-case1} and \ref{Tab:BR-case2} to those for fermionic final states. Two cases are considered: in Table~\ref{Tab:BR-case1}, a $W'$ with mixing $\xi_{W\text{-}W^\prime}=(M_W/M_{W'})^2$, as suggested by Eq.~(\ref{Eq:define-xi} with ${\cal C}=1$, and in Table~\ref{Tab:BR-case2}, a $W'$ with mixing $\xi_{W\text{-}W^\prime}=10^{-3}$. 
In evaluation of the diboson decay partial widths of $W'$ in both cases, the relation of $\Gamma_{W'}^{WH}=\Gamma_{W'}^{WZ}$ is assumed.
The presence of the two last diboson decay channels is due to $W$-$W^\prime$ mixing and is often neglected, however for large $W'$ masses there is an enhancement that cancels the suppression due to the mixing leading to a linear increase of the diboson partial widths with $M_{W'}$. This is in contrast to the second case where
for a fixed (mass-independent) value of $\xi_{W\text{-}W^\prime}$, the diboson branching ratio is seen to grow rapidly with mass, in fact as the fifth power \cite{Altarelli:1989ff}. This feature of the model allows for a high sensitivity.

%%%%%%%%%%%%%%%%%%%%%%%%%%%%%%%%%%%%%%%%%%%%%%%%%
\begin{table}[htb]
\caption{$W^\prime$ branching ratios in per cent for
$\xi_{W\text{-}W^{\prime}}=10^{-3}$.}
\begin{center}
\begin{tabular}{|l|c|c|c|c|c|}
\hline\hline
\ \ \ \ $M_{W'}$~[TeV]  & 1 & 2 & 3 & 4 & 5 \\
\hline
 BR $\to$ light quarks
 & 50.5 & 49.7 & 47.7 & 43.4 & 36.4 \\
\hline
BR $\to t\bar b$
&   24.1 & 24.5 & 23.8 & 21.6 & 18.2 \\
\hline
BR $\to \ell\nu$
& 25.3 & 24.8 & 23.9 & 21.7 & 18.2 \\
\hline
BR $\to WZ+WH$
& 0.1 & 1.0 & 4.6 & 13.3 & 27.2 \\
\hline
\end{tabular}
\end{center}
\label{Tab:BR-case2}
\end{table}
%%%%%%%%%%%%%%%%%%%%%%%%%%%%%%%%%%%%%%%%%%%%%%%%%

%%%%%%%%%%%%%%%%%%%%%%%%%%%%%%%%%%%%%%%%%%%%%%%%%
\subsection{$Z'\to WW$}
%%%%%%%%%%%%%%%%%%%%%%%%%%%%%%%%%%%%%%%%%%%%%%%%%

In analogy with the $W'$ case, in the calculation of the total width $\Gamma_{Z'}$ we included
$Z'\to f\bar f$, $W^+W^-$, and $ZH$ \cite{Bobovnikov:2018fwt,Barger:2009xg}.
We shall again ignore the couplings of the $Z'$ to any beyond-SM particles such as
right-handed neutrinos, SUSY partners or exotic fermions in
the theory, which may increase the width of the $Z'$ and hence lower
the branching ratio into a pair of $W^\pm$ by the same factor.
The total width $\Gamma_{Z'}$ of the $Z'$ boson can then be written as:
\begin{equation}\label{gamma2}
\Gamma_{Z'} = \sum_f \Gamma_{Z'}^{ff} + \Gamma_{Z'}^{WW} +
\Gamma_{Z'}^{ZH}.
\end{equation}

Similar to the total decay width of the $W'$ boson defined in Eq.~(\ref{gamma}),
the presence of the two last decay channels is due to $Z$-$Z'$
mixing.  Note, that the widths of these two bosonic modes $W^+W^-$
and $ZH$ do not depend on unknown masses of the final states.
For the range of $M_{Z'}$
values  below $\sim 3$ TeV,  the dependence of
$\Gamma_{Z'}$ on the values of the mixing parameter $\xi_{Z\text{-}Z^\prime}$
\cite{Altarelli:1989ff} (within its allowed range)
induced by $\Gamma_{Z'}^{WW}$ and
 $\Gamma_{Z'}^{ZH}$ is unimportant.
Therefore, in this mass range, one can
 approximate the total width as $\Gamma_{Z'} \approx \sum_f
 \Gamma_{Z'}^{ff}$, where the sum runs over SM fermions only.
 In this mass range, the ratio of $\Gamma_{Z'}/M_{Z'}=0.03$ for the EGM from which
 one can appreciate the narrowness of the $Z'$ pole.

%%%%%%%%%%%%%%%%%%%%%%%%%%%%%%%%%%%%%%%%%%%%%%%%%%%%%%%
\begin{figure}[htb]
\refstepcounter{figure} \label{BR-zprime}
\addtocounter{figure}{-1}
\begin{center}
\includegraphics[scale=0.6]{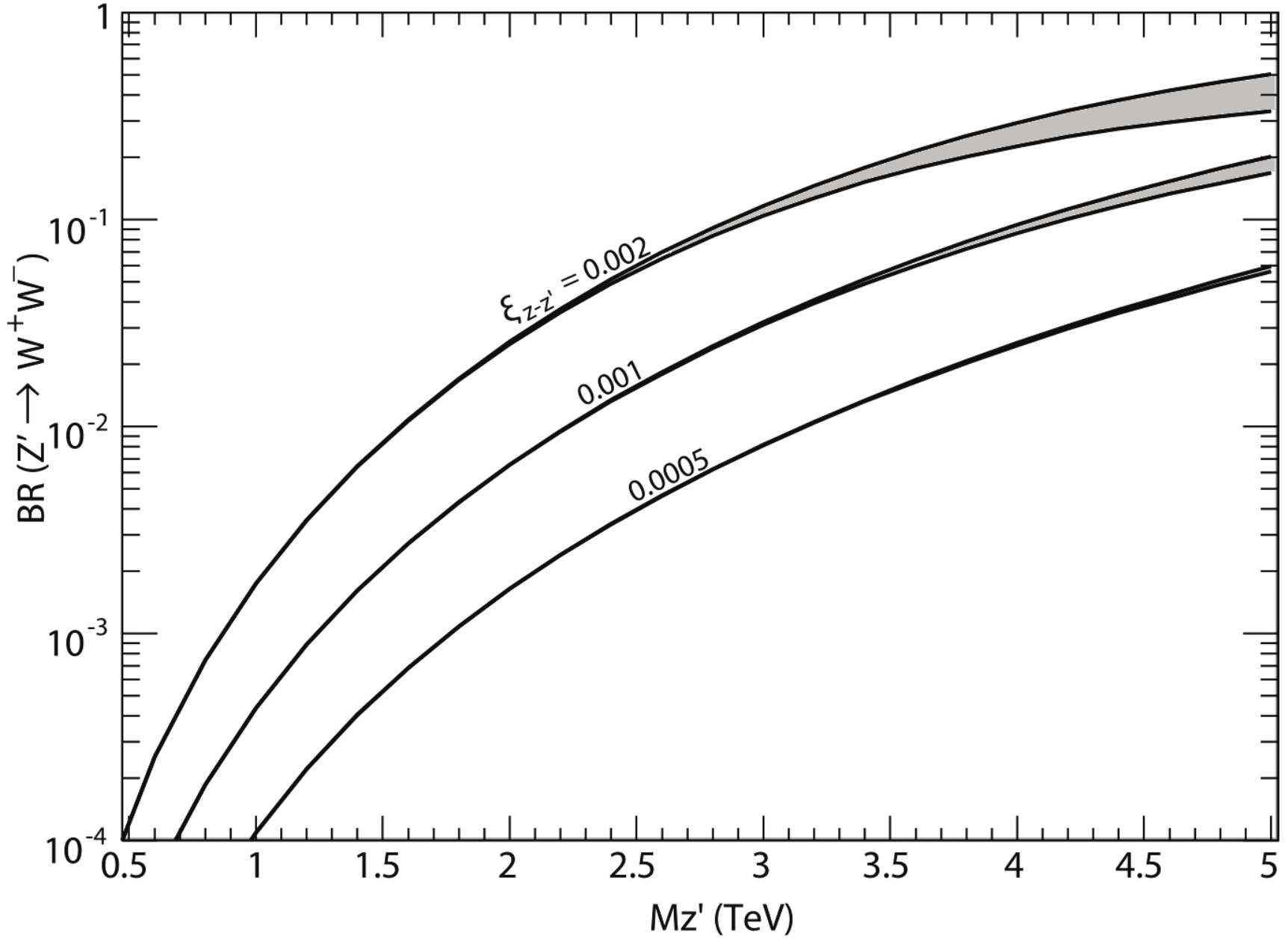}
\end{center}
\caption{Branching fraction $\text{BR}(Z'\to W^+W^-)$ vs.\
$M_{Z'}$ in the EGM for non-zero $Z$-$Z'$ mixing factor
$\xi_{Z\text{-}Z^\prime}=2\cdot 10^{-3}$, $1\cdot 10^{-3}$ and
$5\cdot 10^{-4}$. The shaded bands represent the uncertainty
resulting from the inclusion of the $ZH$ decay mode to the total
decay width $\Gamma_{Z'}$, the upper and lower bounds correspond
to the assumptions $\Gamma_{Z'}^{ZH}=0$ and
$\Gamma_{Z'}^{ZH}=\Gamma_{Z'}^{W^+W^-}$, respectively.}
\end{figure}
%%%%%%%%%%%%%%%%%%%%%%%%%%%%%%%%%%%%%%%%%%%%%%%%%%%%%%%%

However, for large $Z'$ masses, $M_{Z'}>3-5$ TeV, there is an
enhancement that cancels the suppression due to the tiny $Z$-$Z'$
mixing parameter $\xi_{Z\text{-}Z^\prime}$ \cite{Bobovnikov:2018fwt}. While the
``Equivalence theorem'' \cite{Chanowitz:1985hj} might suggest a
value for $\text{BR}(Z'\to ZH)$ comparable to $\text{BR}(Z'\to
W^+W^-)$ up to electroweak symmetry breaking
effects and phase-space factors,  the $Z'ZH$ coupling is quite
model dependent \cite{Barger:1987xw,Barger:2009xg}. We again take an
approach as model-independent as possible, and show our results for two scenarios,
analogous to the corresponding ones for the $W'$ case.
In the first scenario, we treat the model as effectively having a
suppressed  partial width of $Z'\to ZH$ with respect to that of
$Z'\to W^+W^-$, i.e. $\Gamma_{Z'}^{ZH}\ll\Gamma_{Z'}^{WW}$, so
that one can ignore the former.
In this case, numerical
results with our treatment will serve as an upper bound on the
size of the signal. The second scenario concerns the situation when
both partial widths are comparable, $\Gamma_{Z'}^{ZH}\simeq
\Gamma_{Z'}^{WW}$ for heavy $M_{Z'}$
\cite{Barger:1987xw,Barger:2009xg,Dib:1987ur}.

For a fixed mixing factor $\xi_{Z\text{-}Z^\prime}$ and at large $M_{Z'}$ where
$\Gamma_{Z'}^{WW}$ dominates over $\sum_f \Gamma_{Z'}^{ff}$ (assuming
partial width of $\Gamma_{Z'}^{ZH}=0$)
 the total width increases
rapidly with the mass $M_{Z'}$ because of the quintic
dependence on the $Z'$ mass of the $W^+W^-$ mode \cite{Altarelli:1989ff,Bobovnikov:2018fwt}. In this case, the
$W^+W^-$ mode becomes dominant and $\text{BR}(Z' \to W^+W^-)\to
1$, while the fermionic decay channels ($\Gamma_{Z'}^{ff}\propto M_{Z'}$) are increasingly
subdominant.

For the EGM, the $Z$-$Z^\prime$ mixing parameter $\xi_{Z\text{-}Z^\prime}$ is constrained at
the level of a few per mil \cite{Erler:2009jh} from an analysis of the $Z^\prime$ model against available electroweak precision data, resulting in  $\xi_{Z\text{-}Z^\prime}^{\rm EW}<2.6\cdot 10^{-3}$.
 In Fig.~\ref{BR-zprime}  we plot
$\text{BR}(Z'\to W^+W^-)$ vs $M_{Z'}$ for the EGM  and mixing factor
$\xi_{Z\text{-}Z^\prime}$  ranging from 0.0005 to 0.002. The case when
$\Gamma_{Z'}^{ZH}=\Gamma_{Z'}^{W^+W^-}$ is also shown in Fig.~\ref{BR-zprime}.

It should be stressed that the boost of the branching ratio for high values of $M_{W'}$ and $M_{Z'}$, illustrated in Figs.~\ref{BR-wprime} and \ref{BR-zprime}, plays an important role in the following analysis.

%%%%%%%%%%%%%%%%%%%%%%%%%%%%%%%%%%%%%%%%%%%%%%%%%%%%%%%%%%%%
\section{Analysis}
\label{sect:analysis}
\setcounter{equation}{0}
%%%%%%%%%%%%%%%%%%%%%%%%%%%%%%%%%%%%%%%%%%%%%%%%%%%%%%%%%%%%

%%%%%%%%%%%%%%%%%%%%%%%%%%%%%%%%%%%%%%%%%%%%%%%%%
\subsection{Production and decay of $W'\to WZ$}
\label{WprimeWZ}
%%%%%%%%%%%%%%%%%%%%%%%%%%%%%%%%%%%%%%%%%%%%%%%%%

Here, we present an analysis, employing the most recent measurements of
diboson processes provided by ATLAS \cite{Aad:2019fbh}
with the full Run~2 data set with time-integrated luminosity of 139 fb$^{-1}$ as well as, for the sake of comparison,  with a partial Run~2 data set with time integrated luminosity of 36.7 fb$^{-1}$ \cite{Aaboud:2017eta}.  As mentioned above, ATLAS analyzed the $WZ$ production in the process (\ref{procWZ}) through the fully hadronic ($qqqq$) final states.\footnote{For the experimental data, ``$qqqq$'' refers to four-jet final states (including gluons).}
In Fig.~\ref{sigma-wprime}, we show the observed
$95\%$ C.L. upper limits on the production cross section times the branching
fraction, $\sigma_{95\%}\times \text{BR}(W'\to WZ)$, as a function
of the $W'$ mass, $M_{W'}$.
The data analyzed comprises $pp$
collisions at $\sqrt{s}=13$ TeV, recorded by the ATLAS (36.7 fb$^{-1}$ and 139
fb$^{-1}$) detector \cite{Aaboud:2017eta,Aad:2019fbh} at the LHC.

 %%%%%%%%%%%%%%%%%%%%%%%%%%%%%%%%%%%%%%%%%%%%%%%%%%%%%%%%
\begin{figure}[htb]
\refstepcounter{figure} \label{sigma-wprime}
\addtocounter{figure}{-1}
\begin{center}
\includegraphics[scale=0.6]{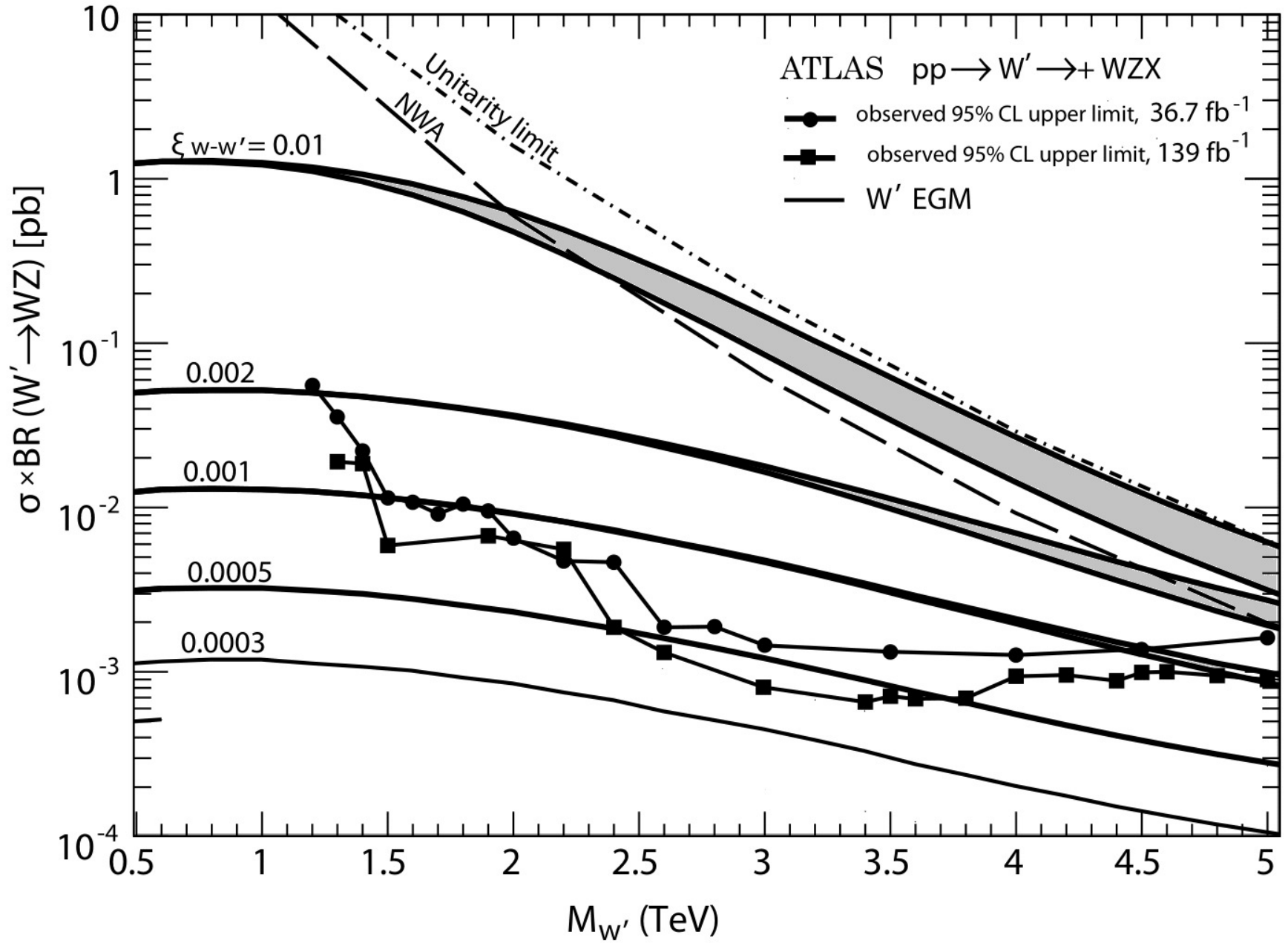}
\end{center}
 \caption{Observed  $95\%$ C.L. upper
limits on the production cross section times the branching
fraction, $\sigma_{95\%}\times \text{BR}(W'\to WZ)$, as a function
of the $W'$ mass, $M_{W'}$, showing ATLAS data on the fully
hadronic final states for $36.7~\text{fb}^{-1}$ \cite{Aaboud:2017eta}
and $139~\text{fb}^{-1}$  \cite{Aad:2019fbh}.
The theoretical production cross sections $\sigma(pp\to W'+X)\times \text{BR}(W'\to
WZ)$ for the ${\rm EGM}$ are calculated from PYTHIA with a $W'$
boson mass-dependent $K$-factor used to correct for NNLO QCD effects,
and given by solid curves, for mixing factor $\xi_{W\text{-}W^\prime}$
ranging from $10^{-2}$ and down to $3\cdot 10^{-4}$. The shaded
bands are defined like in Fig.~\ref{BR-wprime}. The area lying below
the long-dashed curve labelled NWA corresponds to the region
where the narrow-resonance assumption is
satisfied. The lower boundary of the region excluded by the
unitarity violation arguments is indicated by the dot-dashed
curve \cite{Alves:2009aa, Serenkova:2019zav}.}
\end{figure}
%%%%%%%%%%%%%%%%%%%%%%%%%%%%%%%%%%%%%%%%%%%%%%%%%%%%%%%%

%%%%%%%%%%%%%%%%%%%%%%%%%%%%%%%%%%%%%%%%%%%%%%%%%%%%%%%%
\begin{figure}[htb]
\begin{center}
\includegraphics[scale=0.6]{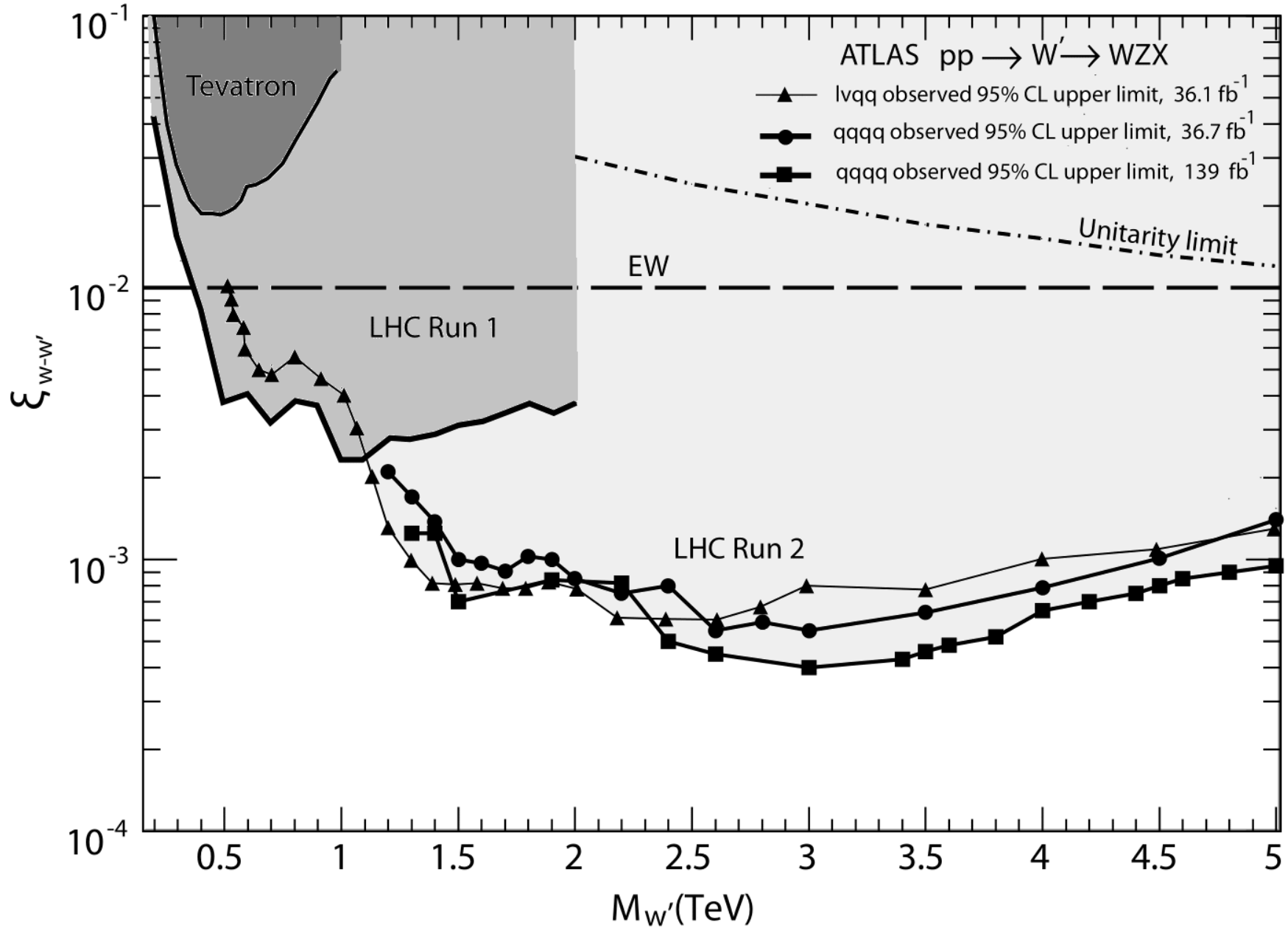}
\end{center}
\caption{95\% C.L. exclusion regions  in the two-dimensional
($M_{W'}$, $\xi_{W\text{-}W^\prime}$) plane obtained from the
precision electroweak data (horizontal dashed straight line
labeled ``EW''), the direct search constraints at the Tevatron in
$p\bar{p}\to WZX$ (the dark shaded area) as well as from the LHC
measurement of $p{p}\to WZX$ at 7 TeV and 8 TeV (Run~1) (the gray
area) and at 13~TeV from diboson $W'\to WZ$ production in hadronic
final states using the partial and full Run~2 ATLAS data set.
Limits obtained from the semileptonic channel $\ell\nu qq$ at
time-integrated luminosity of $36.1~\text{fb}^{-1}$
\cite{Serenkova:2019zav} are overlaid  for comparison. The
combined exclusion region for  the  EGM $W'$ boson obtained  after
incorporating   direct  search  constraints  from the  LHC  Run~2
data set is shown as the light shaded area. The uninarity
limit is shown as a dot-dashed curve.}  \label{wprime-bounds}
\end{figure}
%%%%%%%%%%%%%%%%%%%%%%%%%%%%%%%%%%%%%%%%%%%%%%%%%%%%%%%%

Then, for $W'$ we compute the LHC theoretical production cross
section multiplied by the branching ratio into  $WZ$ bosons,
$\sigma (pp\to W'X)\times {\rm BR}(W'\to WZ)$, as a function of the
two parameters ($M_{W'}$, $\xi_{W\text{-}W^\prime}$) \cite{Serenkova:2019zav}, and compare it with the limits
established by the ATLAS experiment, $\sigma_{95\%} \times {\rm BR}(W'\to WZ)$.
The simulation of signals
for the EGM $W'$ is based on an adapted version of
the leading order (LO) PYHTHIA 8.2
event generator \cite{Sjostrand:2014zea}.
A mass-dependent $K$ factor is adopted to rescale the LO
PYTHIA prediction to the next-to-next-to-leading-order (NNLO) in
$\alpha_s$. The theoretical $W'$ production cross section $\sigma(pp\to
W'X)$ is scaled to an NNLO calculation in
$\alpha_s$ by ZWPROD \cite{Hamberg:1990np}, given by solid curves, and shown
in Fig.~\ref{sigma-wprime} for a mixing factor $\xi_{W\text{-}W^\prime}$ ranging from
$10^{-2}$ and down to $3\cdot 10^{-4}$. The
factorization and renormalization scales are set to the $W'$
resonance mass.

As was explained in connection with Fig.~\ref{BR-wprime}, the upper (lower) boundary
of the  shaded  areas correspond to a
scenario where the contribution of the decay
channel $W'\to WH$ to the total $W'$ decay width of
Eq.~(\ref{gamma}) is $\Gamma_{W'}^{WH}=0$
($\Gamma_{W'}^{WH}=\Gamma_{W'}^{WZ}$). The area below the
long-dashed curve labelled ``NWA'' corresponds to the region where
the $W'$ resonance width is predicted to be less than 5\% of its
mass, corresponding to the best detector resolution of
the searches, where the narrow-resonance assumption is satisfied.
We also show a curve labelled  ``Unitarity limit'' that corresponds to the unitarity bound (see, e.g. \cite{Alves:2009aa} and references therein). In that paper,  it was shown that the saturation of
unitarity in the elastic scattering $W^\pm Z\to W^\pm Z$   leads to
the constraint  ${g_{W'WZ}}_\text{max}=g_{WWZ}\cdot M_Z^2/(\sqrt{3}\,M_{W'}\,M_W)$)
that was exploited in plotting the unitarity bound. This constraint
was obtained under the assumption  that the couplings of the $W^\prime$ to
quarks and to gauge bosons have the same Lorentz structure
as those of the SM but with rescaled strength.

The theoretical curves for the cross sections $\sigma(pp\to W'X)
\times {\rm BR}(W'\to WZ)$, in descending order, correspond to
values of the $W$-$W'$ mixing factor $\xi_{W\text{-}W^\prime}$ from 0.01 to 0.0003.
The intersection points of the measured upper
limits on the production cross section with these theoretical
cross sections for various values of $\xi_{W\text{-}W^\prime}$ give the corresponding lower
bounds on ($M_{W'}$, $\xi_{W\text{-}W^\prime}$), to be displayed  below, in
Fig.~\ref{wprime-bounds}.

The limits arising from the diboson channel are basically excluding
large values of $\xi_{W\text{-}W^\prime}$, strongest at intermediate masses
$M_{W'}\sim 2-4~\text{TeV}$, as illustrated in Fig.~\ref{wprime-bounds}.
Interestingly, Fig.~\ref{wprime-bounds} shows that at moderate and high $W'$ masses, the
limits on $\xi_{W\text{-}W^\prime}$ obtained from the ATLAS diboson resonance
production search at 13 TeV and at time-integrated luminosity of
 139 fb$^{-1}$ are substantially stronger than those derived from the
 low-energy electroweak data, which are of the order $\sim 10^{-2}$ \cite{Tanabashi:2018oca}, as well as those obtained  from the partial ATLAS Run~2 data set with time integrated luminosity of 36.7 fb$^{-1}$   \cite{Aaboud:2017eta}  in the fully hadronic final states, and as well as those obtained  in the semileptonic final state at 36.1 fb$^{-1}$ \cite{Serenkova:2019zav}.

 Comparison of sensitivities to $W'$ of the process (\ref{procWZ}) with different decay channels, e.g., $VV\to\ell\nu qq$ and $qqqq$, can be performed by the matching of $95\%$ C.L. upper limits on the production cross section times the branching fraction, $\sigma_{95\%}\times \text{BR}(W'\to WZ)$, which includes the SM branching fractions of the electroweak bosons to the final states in the analysis channel, effects from detector acceptance, as well as reconstruction and selection efficiencies. ATLAS bounds were included acccording to HEPdata \cite{hepdata}.
 From a comparison of the upper limits on the production cross section times the branching
fraction for semileptonic $\ell\nu qq$ vs.\ fully hadronic $qqqq$ decay channels at 36.1 fb$^{-1}$  and 36.7 fb$^{-1}$, respectively, one can  conclude that the $qqqq$ channel dominates the sensitivity in the higher resonance  mass range  (2.6 TeV $\leq M_{W'} \leq$ 5 TeV), while at lower masses the sensitivity of the semileptonic channel dominates over the fully hadronic one. These features are illustrated in Fig.~\ref{wprime-bounds}.

For reference, we display limits on the $W'$ parameters
from the Tevatron (CDF and D0) as well as from ATLAS and CMS
obtained at 7  and 8 TeV of the LHC data taking in Run~1 denoted as ``LHC Run~1'' \cite{Serenkova:2019zav}.
Fig.~\ref{wprime-bounds} shows that the experiments CDF and D0 at the Tevatron exclude EGM $W'$ bosons with $\xi_{W\text{-}W^\prime}\gsim 2\cdot 10^{-2}$ in the resonance mass range 0.25~TeV $<M_{W'}<$ 1~TeV at the $95\%$ C.L., whereas LHC in Run~1 improved those constraints, excluding  $W'$ boson parameters at $\xi_{W\text{-}W^\prime}\gsim 2\cdot 10^{-3}$ in the mass range 0.2~TeV  $<M_{W'}<$ 2~TeV.

As expected, the increase of the time-integrated luminosity up to
139~fb$^{-1}$ leads to dominant sensitivity of the $qqqq$ channel over the whole resonance mass range of 1.3~TeV $<M_{W'}<$ 5~TeV and it allows to set stronger constraints on the mixing
angle $\xi_{W\text{-}W^\prime}$, resulting in $\xi_{W\text{-}W^\prime} >  4.3\cdot 10^{-4}$  as shown in Fig.~\ref{wprime-bounds}.
Our results extend the sensitivity beyond the corresponding
CDF Tevatron  results \cite{Aaltonen:2010ws}
 as well as the ATLAS and CMS sensitivity attained
at 7 and 8~TeV. Also, for the first time, we set $W'$ limits as functions
of the mass $M_{W'}$ and mixing factor $\xi_{W\text{-}W^\prime}$ at the LHC at 13~TeV
with the partial ATLAS Run~2 data set at time-integrated luminosity of 36.7 fb$^{-1}$ \cite{Aaboud:2017eta},  and with the full ATLAS Run~2 data set with a time-integrated luminosity of 139~fb$^{-1}$.
The exclusion region obtained in this way on the
parameter space of the $W'$ from the full Run~2
data set supersedes the corresponding exclusion area obtained at the LHC at $\sqrt{s}=13$ TeV and
time-integrated luminosity of 36.1~fb$^{-1}$ in the semileptonic channel as reported in \cite{Serenkova:2019zav}.
The limits on $W'$ parameters presented in this section obtained from the diboson $WZ$ production in hadronic final states using the full Run~2 ATLAS data set, corresponding to  a time-integrated luminosity of 139~fb$^{-1}$ are the best to date.

%%%%%%%%%%%%%%%%%%%%%%%%%%%%%%%%%%%%%%%%%%%%%%%%%
\subsection{Production and decay of $Z'\to WW$}
%%%%%%%%%%%%%%%%%%%%%%%%%%%%%%%%%%%%%%%%%%%%%%%%%

For the $Z'$ case, the analysis proceeds in a similar fashion.
We show in Fig.~\ref{sigma-zprime} the observed  $95\%$
C.L. upper limits on the production cross section times the
branching fraction, $\sigma_{95\%}\times \text{BR}(Z'\to
W^+W^-)$, as a function of the $Z'$ mass, $M_{Z'}$.
Then, for $Z'$ we compute the LHC production cross section multiplied
by the branching ratio into two $W$ bosons, $\sigma \times {\rm
BR}(Z'\to W^+ W^-)_{\rm theory}$, as a function of the two parameters
($M_{Z'}$, $\xi_{Z\text{-}Z^\prime}$), and compare it with the limits established by the
ATLAS experiment, $\sigma_{95\%} \times {\rm BR}(Z'\to W^+ W^-)$.
Our strategy in the present analysis is to adopt the SM backgrounds that have been carefully
evaluated by the experimental collaborations and contained in $\sigma_{95\%} \times {\rm BR}(Z'\to W^+ W^-)$ and simulate only the $Z'$ signal.
Comparison of the $95\%$ C.L. upper limits on the production cross section
times the branching fraction, $\sigma_{95\%}\times \text{BR}(Z'\to
W^+W^-)$, as a function of the $Z'$ mass based on the ATLAS
data of the fully hadronic final states for $36.7~\text{fb}^{-1}$
\cite{Aaboud:2017eta} and $139~\text{fb}^{-1}$ \cite{Aad:2019fbh} demonstrates the dominating sensitivity to $Z'$ of the latter time-integrated luminosity data with respect to the former one, over the whole $Z'$ mass range.
%%%%%%%%%%%%%%%%%%%%%%%%%%%%%%%%%%%%%%%%%%%%%%%%%%%%%%%%
\begin{figure}[hbt]
\refstepcounter{figure} \label{sigma-zprime}
\addtocounter{figure}{-1}
\begin{center}
\includegraphics[scale=0.6]{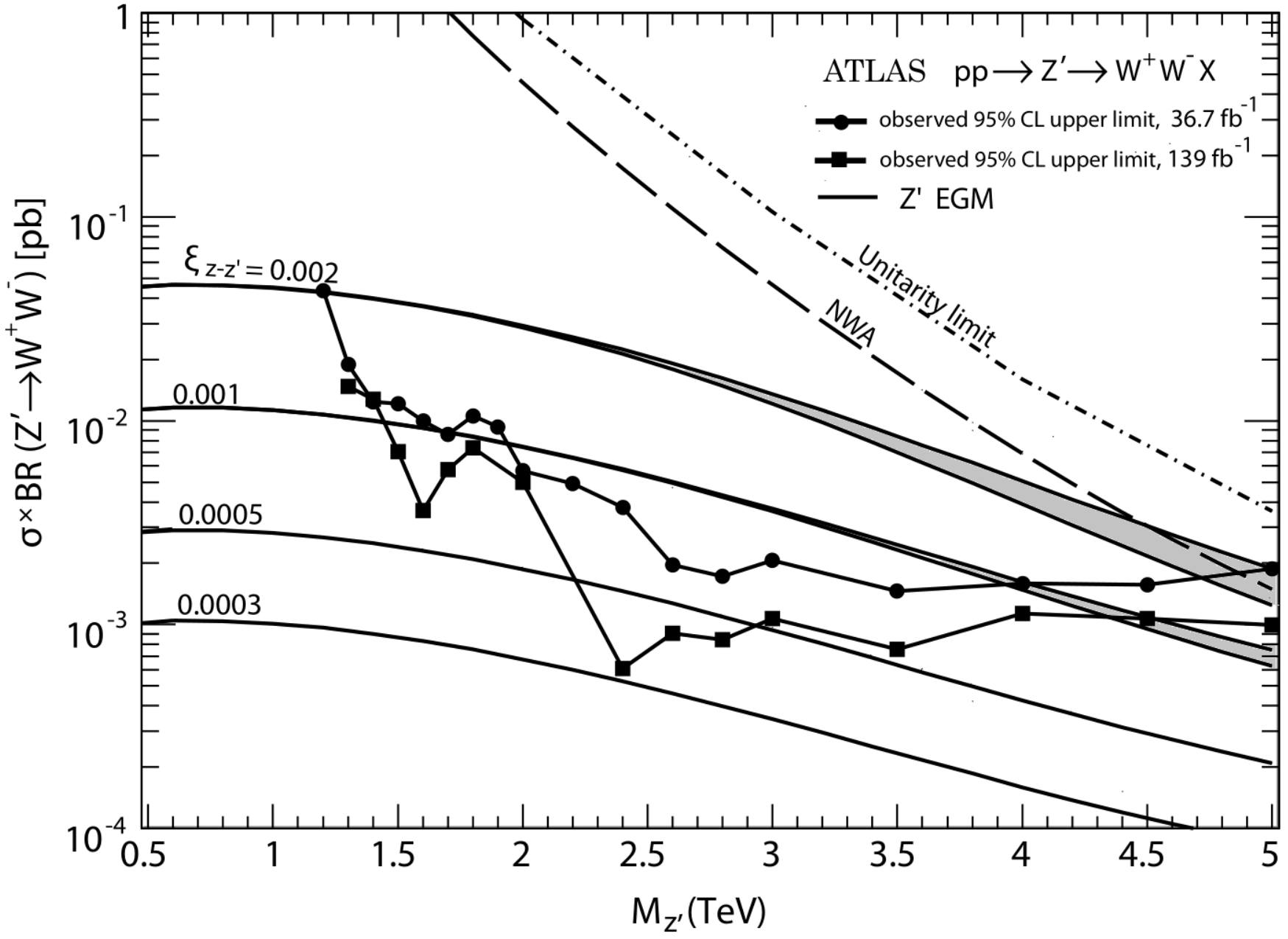}
\end{center}
 \caption{
 Observed  $95\%$ C.L. upper limits on the production cross section
times the branching fraction, $\sigma_{95\%}\times \text{BR}(Z'\to
W^+W^-)$, as a function of the $Z'$ mass, $M_{Z'}$, showing ATLAS
data of the fully hadronic final states for $36.7~\text{fb}^{-1}$
\cite{Aaboud:2017eta} and $139~\text{fb}^{-1}$ \cite{Aad:2019fbh}.
Theoretical production cross sections
$\sigma(pp\to Z'+X)\times \text{BR}(Z'\to W^+W^-)$ for the ${\rm
EGM}$ are calculated from PYTHIA with a $K$-factor used to
correct for NNLO QCD effects, and given by solid curves,
for mixing factors $\xi_{Z\text{-}Z^\prime}$ ranging from $2\cdot 10^{-3}$ and down to
$3\cdot 10^{-4}$. The shaded bands are defined like in
Fig.~\ref{BR-zprime}. The area lying below the long-dashed curve
labelled NWA  corresponds to the
region where the narrow-resonance
assumption is satisfied. The lower boundary of the region excluded
by the unitarity violation arguments is also indicated by the
dot-dashed curve \cite{Alves:2009aa,Bobovnikov:2018fwt}.
}
\end{figure}
%%%%%%%%%%%%%%%%%%%%%%%%%%%%%%%%%%%%%%%%%%%%%%%%%%%%%%%%

%%%%%%%%%%%%%%%%%%%%%%%%%%%%%%%%%%%%%%%%%%%%%%%%%%%%%%%%
\begin{figure}[hbt]
\begin{center}
\includegraphics[scale=0.6]{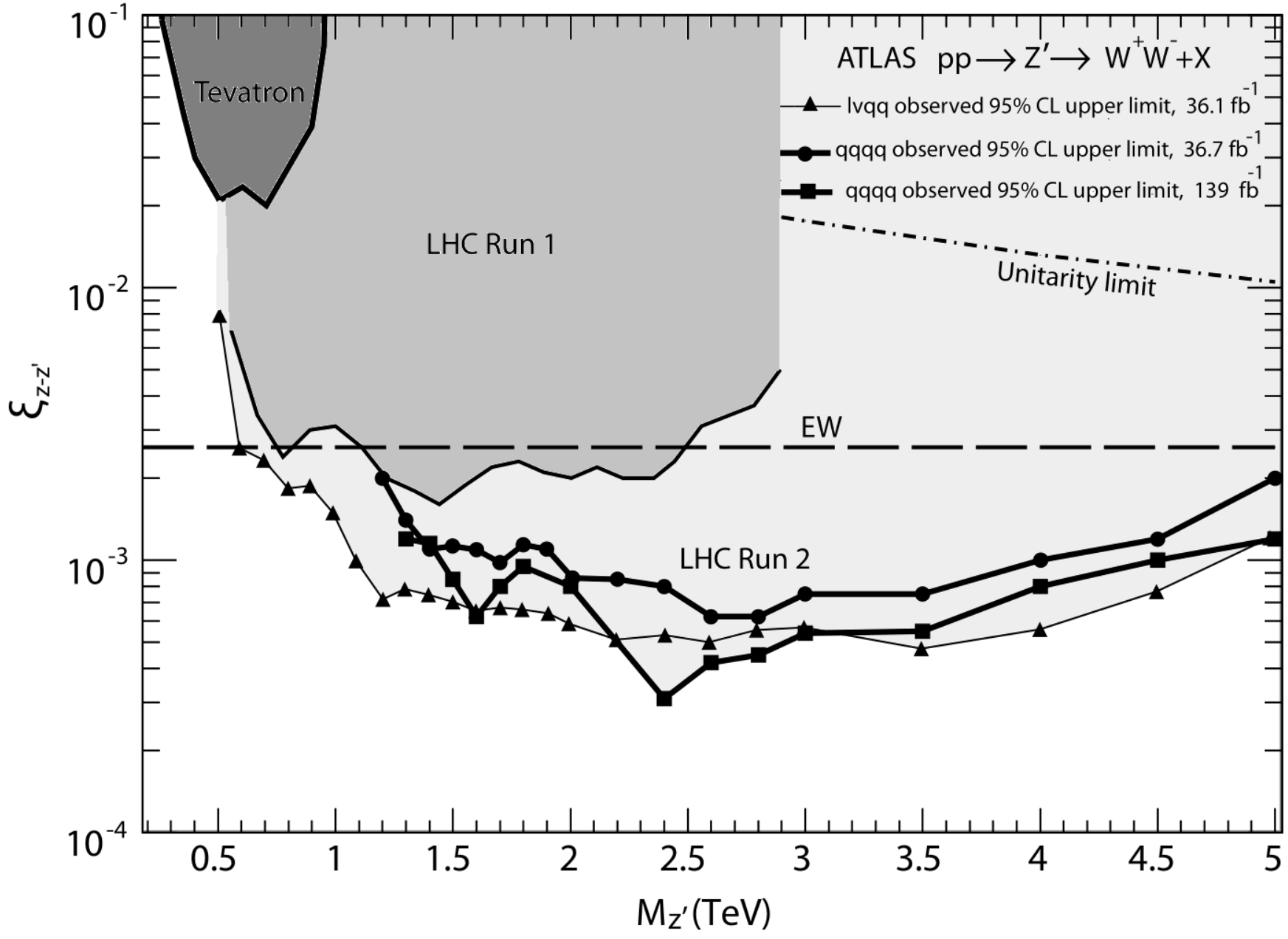}
\end{center}
\caption{ 95\% C.L. exclusion regions in the two-dimensional
($M_{Z'}$, $\xi_{Z\text{-}Z^\prime}$) plane obtained after
incorporating direct search constraints from the CDF and D0
collaborations which are referred to as Tevatron (the dark shaded
area) in $p\bar{p}\to W^+W^-X$  as well as those derived   from
the LHC  measurement of $p{p}\to WWX$ in Run~1 (the gray area)
\cite{CMS:2016wev} and 13 TeV from diboson $Z'\to WW$ production
in hadronic final states using the partial and full Run~2 ATLAS
data set. Also shown is the exclusion from the precision
electroweak (EW) data \cite{Erler:2009jh}. Limits obtained from
the semileptonic channel $\ell\nu qq$ at time-integrated
luminosity of $36.1~\text{fb}^{-1}$ \cite{Bobovnikov:2018fwt} are
overlaid  for comparison. Combined exclusion region for  the  EGM
$Z'$ boson obtained  after  incorporating direct  search
constraints  from the  LHC  Run~2 data set is shown as the light
shaded area.  The uninarity limit is shown as a dot-dashed
curve.} \label{bounds-zprime}
\end{figure}
%%%%%%%%%%%%%%%%%%%%%%%%%%%%%%%%%%%%%%%%%%%%%%%%%%%%%%%%

In Fig.~\ref{sigma-zprime}, the theoretical production cross section
$\sigma\times \text{BR}(Z'\to W^+W^-)_{\rm theory}$ for $Z'$
boson of the EGM, is calculated from PYTHIA 8.2
\cite{Sjostrand:2014zea} adapted for such kind of analysis.
Higher-order QCD corrections to the signal were
estimated using a $K$-factor, for which we adopt a
mass-independent value of 1.9
\cite{Frixione:1993yp,Agarwal:2010sn,Gehrmann:2014fva}. These
theoretical curves for the cross sections, in descending order,
correspond to values of the $Z$-$Z'$ mixing factor $\xi_{Z\text{-}Z^\prime}$ from 0.002
to 0.0003. The intersection points of the expected measured
upper limits on the production cross section with this
theoretical cross section for various values of $\xi_{Z\text{-}Z^\prime}$ give the
corresponding lower bounds on ($M_{Z'}$, $\xi_{Z\text{-}Z^\prime}$), to be presented
in Fig.~\ref{bounds-zprime}.
For reference, we plot also a curve labelled ``Unitarity limit'' that corresponds to the
unitarity bound \cite{Alves:2009aa,Bobovnikov:2018fwt}. In \cite{Alves:2009aa},  it was shown that the saturation of unitarity in the elastic scattering $W^+W^-\to W^+W^-$ leads to
the constraint $g_{Z'WW_{max}}=g_{ZWW}\cdot (M_Z/\sqrt{3}M_{Z'})$
that was exploited in plotting the unitarity bound.

Different bounds on the $Z'$ parameter space are collected in Fig.~\ref{bounds-zprime}, showing that at
high $Z'$ masses, the limits on $\xi_{Z\text{-}Z^\prime}$ obtained from
the full Run~2 data set collected  at $\sqrt{s}=13$ TeV and
recorded by the ATLAS detector
are substantially stronger than those derived from the global analysis
of the precision electroweak data \cite{Erler:2009jh}, which is also displayed. In this Fig.~\ref{bounds-zprime}, we display limits on the $Z'$ parameters in the EGM from the Tevatron exclusion \cite{Aaltonen:2010ws},
as well as those derived from the CMS measurement of $p{p}\to WWX$ in Run~1 \cite{CMS:2016wev}.

Limits obtained from the semileptonic channel $\ell\nu qq$ are also shown for comparison \cite{Bobovnikov:2018fwt}. Below (above) a resonance mass value of about 2.2~TeV (3~TeV), the semileptonic channel at time-integrated luminosity of $36.1~\text{fb}^{-1}$ dominates the sensitivity, while in the resonance mass range 2.2~TeV$\leq M_{Z'}\leq$3~TeV the all-hadronic channel at luminosity of $139~\text{fb}^{-1}$ is most sensitive.
As for a comparison of the sensitivities of different channels, semileptonic vs.\ fully hadronic final states,  at the LHC at 13~TeV with partial ATLAS Run~2 data set, Fig.~\ref{bounds-zprime} shows that the $\ell\nu qq$ channel dominates the sensitivity over the whole resonance mass range 0.5~TeV$\leq M_{Z'}\leq$ 5~TeV.

%%%%%%%%%%%%%%%%%%%%%%%%%%%%%%%%%%%%%%%%%%%%%%%%%
\begin{table}[htb]
\caption{Upper limits  on  mixing parameters $\xi_{W\text{-}W^\prime}$ and $\xi_{Z\text{-}Z^\prime}$ at 95\% C.L.  in the EGM, processes and experiments.}
\begin{center}
\begin{tabular}{|>{\small}c|>{\small}c|>{\small}c|>{\small}c| }
 \hline
collider, process   & $\xi_{W\text{-}W^\prime}$ &   $\xi_{Z\text{-}Z^\prime}$ & @$ M_{V'}$
 \\
   &  &  &  (TeV) \\
\hline
Tevatron, $ p\bar{p} \to W'/Z'\to WZ/WW\,(\to l\nu\,qq)$ \cite{Aaltonen:2010ws}
&   2$\cdot 10^{-2}$  & 2$\cdot 10^{-2}$ &  0.4--0.9 \\
\hline
electroweak (EW) data \cite{Erler:2009jh, Tanabashi:2018oca}  &
 $\sim 10^{-2}$  &  2.6$\cdot 10^{-3}$ & ...
\\
\hline\hline
LHC@13~TeV, Run~2 &&& \\
 $p{p} \to  W'/Z'\to WZ/WW\,(\to l\nu\,qq)$,  $\Lumint=36.1$ fb$^{-1}$
\cite{Serenkova:2019zav,Bobovnikov:2018fwt}
& $6.0\cdot 10^{-4}$ & $4.7\cdot 10^{-4}$  &  0.5--5.0
 \\ \hline
 $p{p} \to W'/Z'\to WZ/WW\,(\to qqqq)$,  $\Lumint=36.7$ fb$^{-1}$   (this work)
 &  $5.5\cdot 10^{-4}$&  $6.2\cdot 10^{-4}$  &  1.2--5.0
 \\ \hline
 $p{p} \to  W'/Z'\to WZ/WW \,(\to qqqq)$,
 $\Lumint=139$ fb$^{-1}$ (this work) &
$4.3\cdot 10^{-4}$
  & $3.1\cdot 10^{-4}$ &  1.3--5.0  \\
\hline
\end{tabular}
\end{center}
\label{Tab:summary}
\end{table}
%%%%%%%%%%%%%%%%%%%%%%%%%%%%%%%%%%%%%%%%%%%%%%%%%

In Table~\ref{Tab:summary}, we collect our limits on the $W'$ and $Z'$ parameters
for the benchmark EGM model. Also shown in Table~\ref{Tab:summary} are the
current limits on the $W$-$W'$ and $Z$-$Z'$ mixing parameters, $\xi_{W\text{-}W^\prime}$ and $\xi_{Z\text{-}Z^\prime}$, from the Tevatron, derived from studies of diboson $WZ$ and $WW$ pair
production. The limits on $\xi_{V\text{-}V'}$ at the Tevatron assume (as does
the present study) that no decay channels into exotic fermions or
superpartners are open to the $W'$ and $Z'$. Otherwise, the limits would
be moderately weaker. Table~\ref{Tab:summary} shows that the limits on $\xi_{V\text{-}V'}$ from the EW
precision data are generally stronger than those from the
preceding Tevatron collider.
The LHC  operating in Run~1 has almost the same sensitivity to mixing parameters as that reached from analysis of low-energy electroweak data.  The only difference is that the Run~1 limits (as analyzed here) depend on the resonance mass ($M_{V'}$) whereas the EW constraints are completely independent.
In addition, the LHC limits obtained in Run~2 at 13~TeV, and time-integrated luminosity, $\Lumint=139$ fb$^{-1}$,
improve the EW limits by a factor of approximately one order, depending on the resonance mass.

%%%%%%%%%%%%%%%%%%%%%%%%%%%%%%%%%%%%%%%%%%%%%%%%%%%%%%%%%%%%
\section{Concluding remarks}
\label{sect:conclusions}
\setcounter{equation}{0}
%%%%%%%%%%%%%%%%%%%%%%%%%%%%%%%%%%%%%%%%%%%%%%%%%%%%%%%%%%%%

Exploration of the diboson $WZ$ and $WW$  production at the LHC with 13~TeV
data set allows to place stringent constraints on the $W$-$W'$  and $Z$-$Z'$ mixing parameters as well as on the $W'$ and $Z'$ masses, respectively. We derived such limits by using the full ATLAS Run~2 data set recorded at the CERN LHC, with integrated luminosity of 139~fb$^{-1}$.

By comparing the experimental limits to
the theoretical predictions for the total cross section of the $W'$ and $Z'$
resonant production and its subsequent decay into $WZ$ or $WW$ pairs, we
show that the  derived constraints on the mixing parameters, $\xi_{W\text{-}W^\prime}$ and $\xi_{Z\text{-}Z^\prime}$,  for
the EGM model, are substantially improved
with respect to those obtained from the global analysis of low
energy electroweak data, as well as from the diboson production study
performed at the Tevatron and those based on the LHC Run~1.

In addition, our work shows that accounting for the contribution of the
$V'$ boson decay channels, $V'\to VH$  (where $V=W/Z$ and $V^{\prime}=W^{\prime}/Z^{\prime}$),  to the
total width $\Gamma_{V'}$ does not dramatically affect the bounds
on the mixing parameter $\xi_{V\text{-}V'}$ obtained in the  scenario of
a vanishing $VH$ mode, $\Gamma_{V'}^{VH}=0$. Namely, it turns out that for
the higher resonance $V'$ masses of our interest the constraints on
$V$-$V'$ mixing are relaxed very little as illustrated in Figs.~\ref{sigma-wprime} and \ref{sigma-zprime} and discussed in Refs.~\cite{Bobovnikov:2018fwt,Serenkova:2019zav}.

We limited ourselves here to an analysis of the full Run 2 ATLAS data set, the corresponding CMS data set is currently unavailable.

%%%%%%%%%%%%%%%%%%%%%%%%%%%%%%%%%%%%%%%%%%%%%%%%%%%%%%%%%%%%
\vspace*{0mm}

\section*{Acknowledgements}
This research has been partially supported by the Abdus Salam ICTP
(TRIL Programme). The work of PO has been supported by the
Research Council of Norway. AAP thanks the CERN TH department for support.

\raggedright
%%%%%%%%%%%%%%%%%%%%%%%%%%%%%%%%%%%%%%%%%%%%%%%%%%%%%%%%%%%%

\end{document}